\documentclass{iau} 
\usepackage{graphicx}

\title[Origin of the Galactic halo] %% give here short title %%
{Origin of the Galactic Halo: \\ accretion vs. in situ formation}

\author[E. Spitoni et al.]   %% give here short author list %%
{Emanuele Spitoni$^1$,
  Fiorenzo Vincenzo{$^2$}, Francesca Matteucci{$^{1,3}$}, \and Donatella Romano{$^4$}}
\affiliation{$^1$Dipartimento di Fisica,  Universit\`a di Trieste\\Via G. Tiepolo 11, 34131 Trieste, Italia \\ email: {\tt spitoni@oats.inaf.it} \\[\affilskip]
  $^2$ University of Hertfordshire, Hatfield, Hertfordshire,\\ Hertfordshire, AL10 9AB, UK\\[\affilskip]
   $^3$ INFN, Sezione di Trieste, via A. Valerio 2, 34127 Trieste, Italy
 \\[\affilskip] 
$^4$ INAF, Area della Ricerca - via Piero Gobetti, 101 - 40129 Bologna - ITALY}

\pubyear{2017}
\volume{334}  %% insert here IAU Symposium No.
\jname{Rediscovering our Galaxy}
\editors{C. Chiappini, I. Minchev, E. Starkenburg, \& M. Valentini, eds.}
\begin{document}

\maketitle

\begin{abstract}

We test the hypothesis that the classical and
ultra-faint dwarf spheroidal satellites of the our Galaxy have been the
building blocks of the Galactic halo by comparing their [O/Fe] and
[Ba/Fe] vs. [Fe/H] patterns with the ones observed in Galactic halo
stars. The  [O/Fe] ratio deviates substantially from the observed
abundance ratios in the Galactic halo stars for [Fe/H] $>$ -2
dex, while they overlap for lower metallicities.
On the other hand, for the neutron capture elements, the discrepancy
is extended at all the metallicities, suggesting that the majority of
stars in the halo are likely to have been formed in situ.
We present the results for a model considering the
effects of an enriched gas stripped from dwarf satellites on the
chemical evolution of the Galactic halo.
We find that the resulting chemical abundances of the halo stars
depend on the adopted infall time-scale, and the presence of a
threshold in the gas for star formation.
%% add here a maximum of 10 keywords, to be taken form the file <Keywords.txt>
\end{abstract}
\firstsection % if your document starts with a section,
              % remove some space above using this command.

\section{Introduction}

The $\Lambda$CDM paradigm predicts that a Milky Way-like galaxy must have
formed by the assemblage  of a large number of smaller systems. In
particular, dwarf spheroidal galaxies (dSphs) were proposed in the
past as the best candidate small progenitor objects, which merged
through cosmic time to eventually give rise to  the stellar Galactic halo  (e.g. Grebel 2005).  On the other hand, Fiorentino et al. (2015) using RR Lyrae stars as
tracers of the Galactic halo (GH) ancient stellar component, showed that
dSphs do not appear to be the major building-blocks of the
GH, estimating  an extreme upper limit of
50\% to their contribution.
\begin{figure}[b]
\centering	  \includegraphics[scale=0.33]{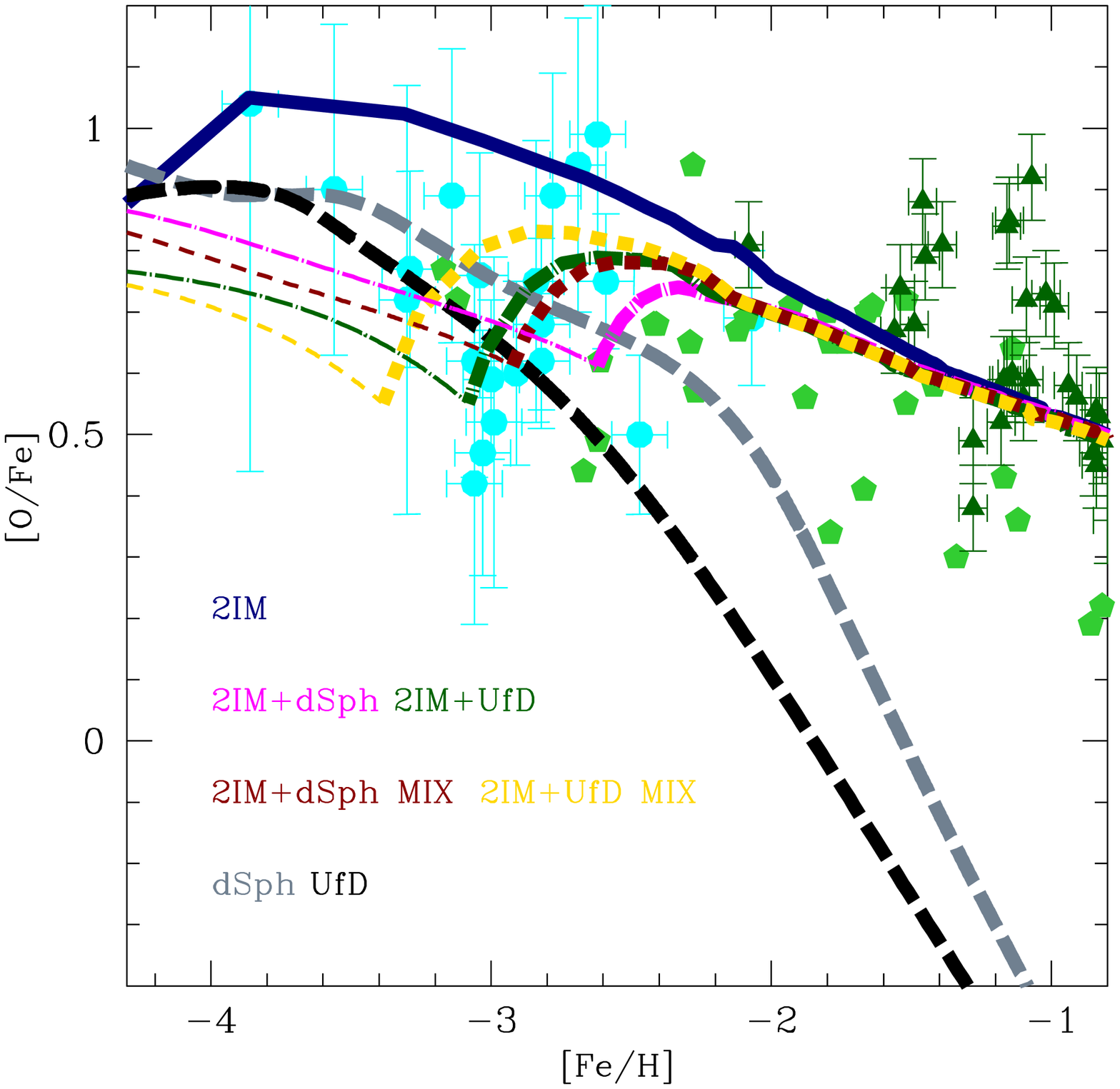}
  \centering \includegraphics[scale=0.33]{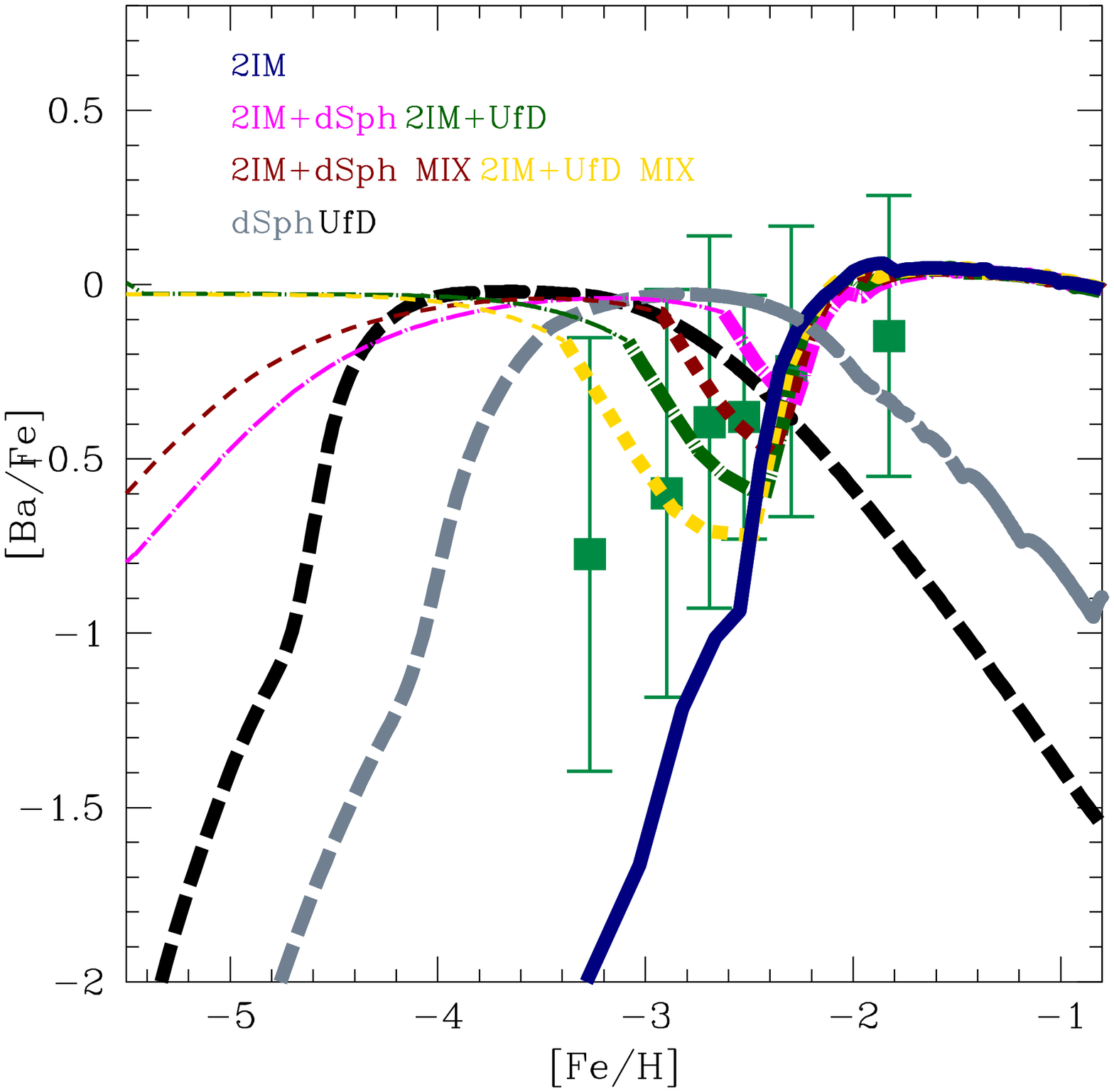}
\caption{[O/Fe] vs [Fe/H]  (left panel) and [Be/Fe] vs [Fe/H]  (right panel) ratios  in the GH  in the solar
neighborhood for the reference model 2IM are drawn with the solid blue
line. {\it Models with the enriched infall from
dSph}: the magenta dashed dotted line and the red short dashed line
represent the models 2IM+dSph and 2IM+dSph MIX, respectively. {\it
Models with the enriched infall from UfDs}: the
green dashed dotted line and the yellow short dashed line represent
the models 2IM+UfD and 2IM+UfD MIX, respectively.   Thinner lines
indicate the ISM chemical evolution phases in which the SFR did not
start yet in the GH, and during which stars are no
created.  {\it Models of the dSph and UfD galaxies}: The long dashed
gray line represents the abundance ratios for the dSph galaxies,
whereas long dashed black line for the UfD galaxies. {\it
Observational O data  of the GH:} Cayrel et al. (2004) (cyan
circles), Akerman et al.  (2004) (light green pentagons), Gratton et
al. (2003) (dark green triangles). 
 {\it Observational  Ba data of the GH:} Frebel  (2010).
 }
\label{O1}
\end{figure}
 \begin{figure}[b]
	 \centering \includegraphics[scale=0.33]{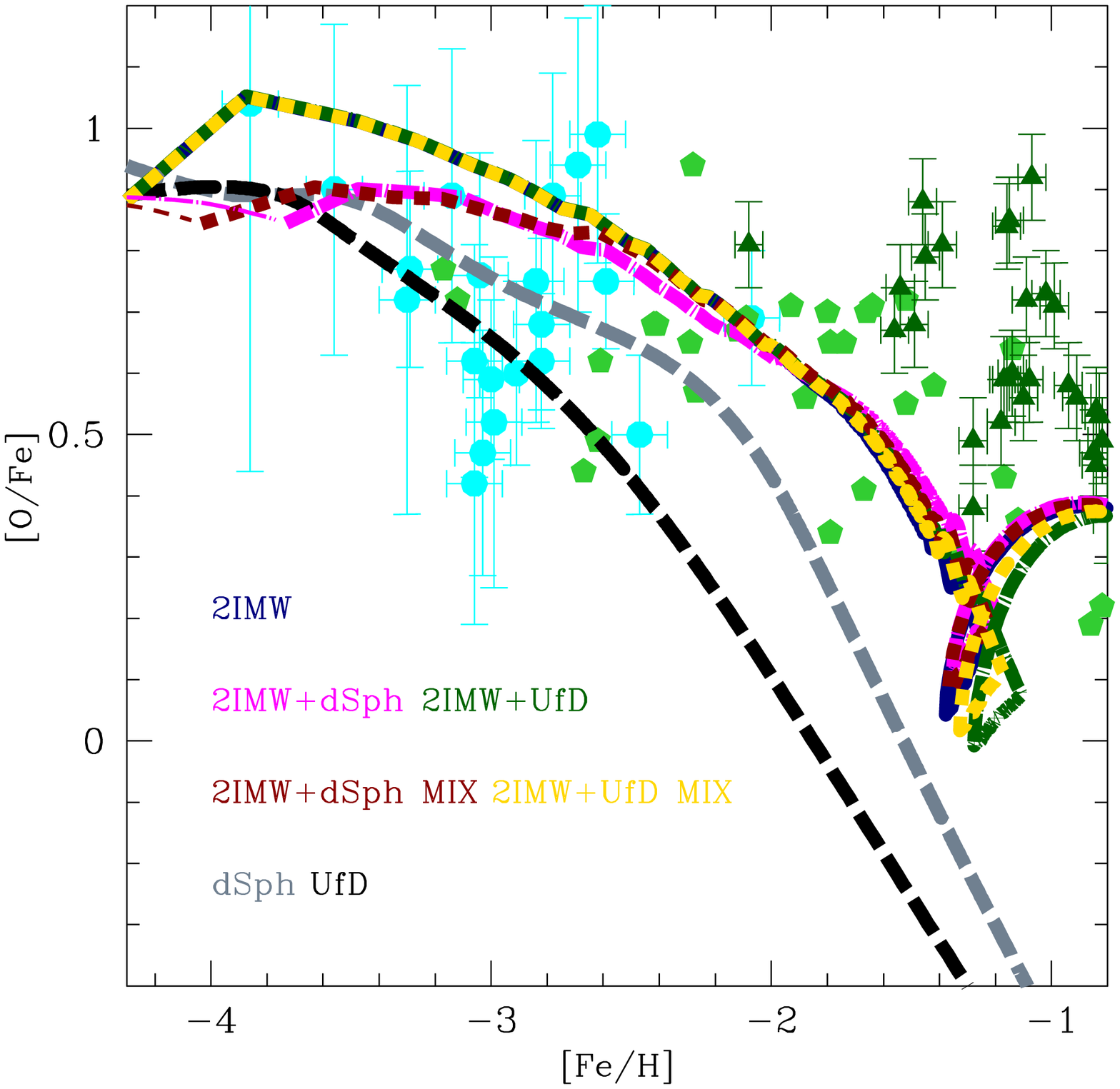}
    \centering \includegraphics[scale=0.33]{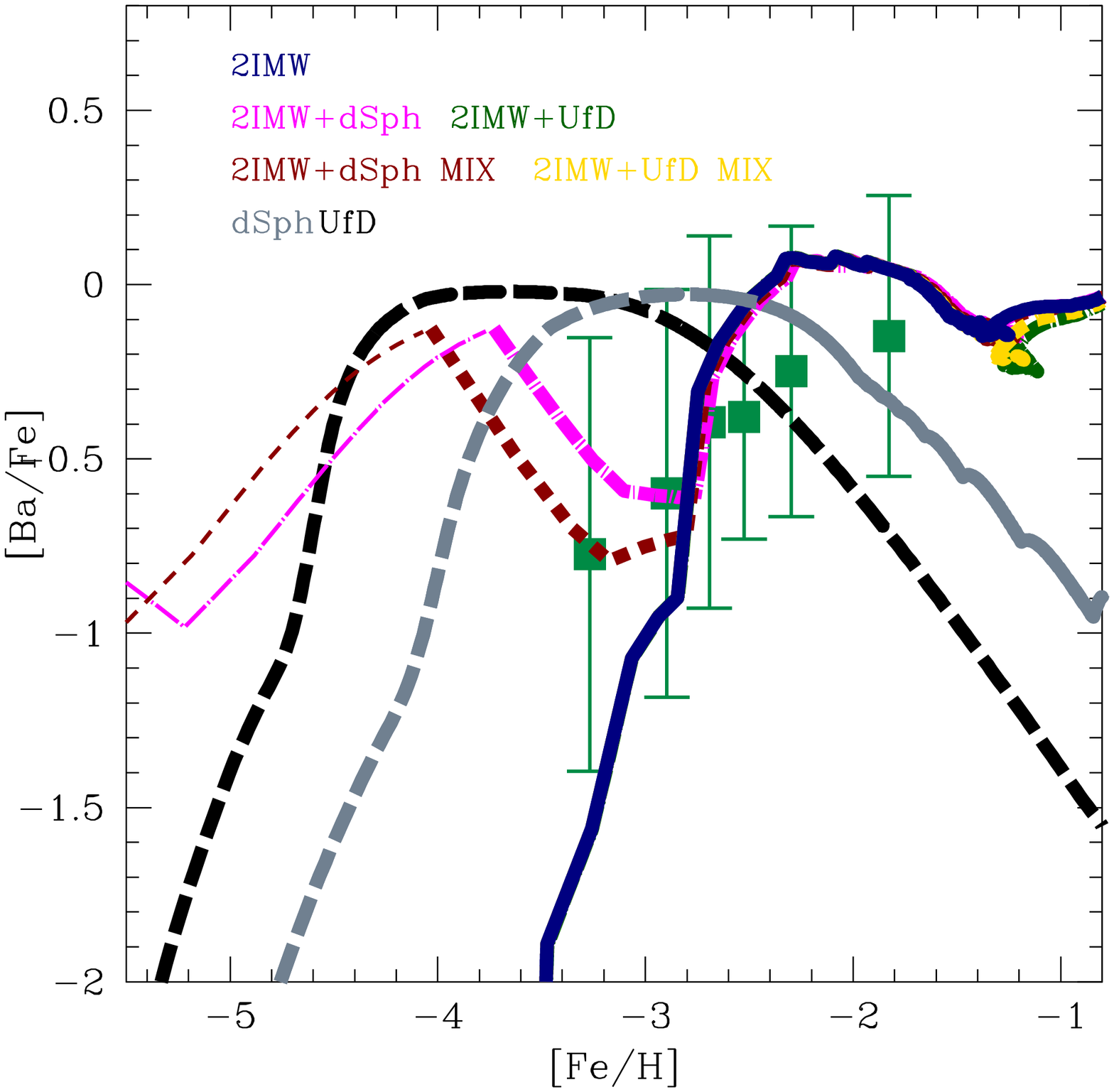}

\caption{ As in Fig. 1  but for the 2IMW model.  
} 
\label{Ba1}
\end{figure}
The SDSS (York et al. 2000) discovered  a new class of objects characterized by extremely low
 luminosities, high dark matter content, and very old and iron-poor
 stellar populations:  the ultra faint dwarf
 spheroidal galaxies (UfDs).
In Spitoni et al. (2016) we test the hypothesis that dSph and UfD
galaxies have been the building blocks of the GH, by
assuming that the halo formed by accretion of stars belonging to these
galaxies. Moreover,  extending  the results of Spitoni (2015) to detailed
chemical evolution models in which the IRA is relaxed, we explored the scenario, in which the 
GH formed by accretion of chemically enriched gas
originating from dSph and UfD galaxies.

\section {The chemical evolution models} 

For the Milky Way we  consider the following two reference chemical evolution models:
\begin{enumerate}
\item The  classical two-infall  model (2IM) presented by Brusadin  et al. (2013). The Galaxy is assumed to have
formed by means of two main infall episodes: the first formed the halo and the thick disk  (with an infall time scale $\tau_H=0.8$ Gyr), the second one the thin disc ($\tau_D=7$ Gyr). 

\item The two-infall model plus outflow of Brusadin et al. (2013; here we  indicate it as the 2IMW model) with  $\tau_H=0.2$ Gyr. In this model  a gas outflow occurring during the GH with a rate proportional to the SFR through a free parameter is considered.  
  \end{enumerate}
In Tables 1, 2, and 3 of Spitoni et al. (2016) all the adopted
parameters of the Milky Way, dSph and UfD models 
are reported.
Here, we only underline that UfDs are characterized by a very
small star formation efficiency (SFE) (0.01 Gyr$^{-1}$) and by an extremely
short time scale of formation (0.001 Gyr). 
The time
at which the galactic wind starts in  dSphs is at 0.013 Gyr after the
galactic formation, whereas for UfDs at 0.088 Gyr.
As expected, the  UfD galaxies develop a  wind at later times  because of the 
smaller adopted SFE.
 The  nucleosynthesis prescriptions are the ones of Romano et al. (2010, model 15) and for Ba, the ones of Cescutti et al. (2006, model 1).

Concerning the model for the GH with the enriched infall we assume that   the gas infall law is
the same as in the 2IM or 2IMW models and it is only considered a time
dependent chemical composition of the infall gas mass.
We tested two different models:

\begin{itemize}

\item Model i): The infall of gas which forms the GH is considered primordial up to the time at which the galactic wind  in dSphs (or UfDs) starts.  After this moment, the infalling gas presents the chemical abundances of the  wind.   In Figs. 1 and 2  we refer to this model with the label "2IM(W)+dSph''or ``2IM(W)+UfD''.

\item Model ii): we explore the case of a diluted infall of for  the GH. In particular, after the galactic wind develops in the dSph (or UfD) galaxy,  
the infalling gas has a chemical composition which, by $50$ per cent,
is contributed by the dSph (or UfD) outflows; the remaining $50$ per
cent is contributed by primordial gas of a different extra-galactic
origin.   In all the successive
figures and in the text, we refer to these models with the labels
``2IM(W)+dSph(UfD) MIX''.  

\end{itemize}

\section{The Results}
\subsection{The Results: the Galactic halo in the model 2IM}
In order to
directly test the hypothesis that GH stars have been
stripped from dSph or UfD systems, in the left panel of Fig. \ref{O1}, the predicted [O/Fe] vs. [Fe/H] abundance
patterns  for  typical dSph and UfD galaxies  are compared with the observed data in GH
stars.  The two models cannot explain the
[O/Fe] plateau which GH stars exhibit for
$\mathrm{[Fe/H]}\ge-2.0$ dex.
 Moreover, in left panel of Fig. 1 we also show the results with the
enriched infall coming from dSph galaxies.  We recall that a key
ingredient of the 2IM model is the presence of a threshold in the gas
density in the star formation (SF) fixed at 4 M$_{\odot}$ pc$^{-2}$ in
the GH. Such a critical threshold is reached only at
$t=0.356\,\mathrm{Gyr}$ from the Galaxy formation.  During the first
0.356 Gyr in both ``2IM+dSph'' and ``2IM+dSph MIX'' models, no stars
are created, and the chemical evolution is traced by the gas
accretion.  After the SF  takes
over, [O/Fe] values increase because of the pollution from massive
stars on short time-scales.  In the ``2IM+dSph'' model the first stars
that are formed have [Fe/H] $>$ -2.4 dex.  In this
case, to explain data for stars with [Fe/H] $<$ -2.4 dex we
need stars formed in dSph systems.  Concerning the results with the
enriched infall from UfD outflow abundances, model results for the GH
still reproduce the data but with the same above mentioned caveat.
In the right panel of Fig. 1, we show
the results for the [Ba/Fe] vs. [Fe/H] abundance diagram.  Chemical
evolution models for dSphs and UfDs fail in reproducing the observed
data, since they predict the [Ba/Fe] ratios to increase at much lower
[Fe/H] abundances than the observed data.  That is due to the very low
SFEs assumed for dSphs and UfDs. The subsequent
decrease of the [Ba/Fe] ratios is due to the large iron content
deposited by Type Ia SNe in the ISM, which happens at still very low
[Fe/H] abundances in dSphs and UfDs.  In the right panel of Fig. 1,
all our models involving an enriched infall from dSphs and UfDs
deviate substantially from the observed trend of the [Ba/Fe]
vs. [Fe/H] abundance pattern in GH stars.
\subsection{The Results: the Galactic halo in the model 2IMW}
In the reference model 2IMW  the SFR starts at 0.05 Gyr. Comparing  model ``2IMW+dSph'' in the left panel of Fig. 2 with  model ``2IM+dSph''
in the left panel of Fig. 1, we can see that the former  shows a   shorter
phase with the enriched infall of gas with SF not yet active than the latter.
The model results for the model ``2IMW+UfD'' in the left panel of Fig. 2 overlap to
the reference model 2IMW at almost all [Fe/H] abundances. 
Since in the UfD galactic model the wind starts at 0.088 Gyr and, at
this instant, in the model 2IMW the SF is already active.
Concerning the [Ba/Fe] vs [Fe/H] ratios (right panel in Fig. 2), we notice that the 2IMW model provides now a better agreement with the
observed data than the 2IM model.
By assuming an enriched infall from
dSph or UfD galaxies, the predicted [Ba/Fe] ratios agree with the
observed data also at $\mathrm{[Fe/H]}<-3\,\mathrm{dex}$.

\section{Conclusions}

\begin{enumerate}

\item The predicted   [O/Fe] vs.[Fe/H] abundance ratios of UfD and dSph chemical evolution models deviate substantially from   the observed data of the GH stars only for [Fe/H] $>$ -2 dex;  we conclude that an evolution in situ in the GH is requested. On the other hand,  we notice that for Ba the
chemical evolution models of dSphs and UfDs fail to reproduce the
observational observed data of the GH stars over the whole
range of [Fe/H].

\item The effects of the enriched infall on the [O/Fe] vs. [Fe/H] plots depend on the infall timescale  of the GH and the presence of a gas threshold in the SF. The most evident effects are present for the model 2IM, characterized by  the longest time scale of formation (0.8 Gyr), and  the  longest period without SF activity  among  all models  presented here.

\item In the presence of an enriched infall of gas  we need  stars produced in dSphs or UfD s  and accreted later to the GH, to explain the data at lowest [Fe/H].

\item The optimal element to test different theories of halo formation is Ba which  is easily measured in low-metallicity stars. In fact, we  have shown  that  the predicted  [Ba/Fe] vs. [Fe/H] relation  in dSphs and UfDs is quite different than in the GH.

\end{enumerate}

\section*{Acknowledgments}
 ES  and FM acknowledge financial support from FRA2016
of the University of Trieste.

\end{document}